\documentclass[a4paper,12pt]{article}

\usepackage{amsmath}
\usepackage{amssymb}
\usepackage{bbm}
\usepackage{graphicx}
\usepackage{accents}
\usepackage{amsthm}
\usepackage{color}
\usepackage{graphics}
\usepackage{epsfig}
\usepackage{cancel}
\usepackage{epstopdf}
\usepackage{hyperref}
\usepackage{xcolor}
\hypersetup{
colorlinks,
linkcolor={blue!50!black},
citecolor={blue!50!black},
urlcolor={blue!80!black}
}
\usepackage{natbib}
\usepackage{stmaryrd} % for double brackets in jump conditions
\usepackage[top = 1in, left = 0.5in, bottom = 1in, right=0.5in]{geometry}

\newcommand{\mbf}[1]{\mathbf{#1}}

\newcommand{\bsym}[1]{\boldsymbol{#1}}

\DeclareMathOperator{\Dgrad}{\bsym{F}}

\usepackage{mdframed} % for framing
\newmdtheoremenv[%
linecolor=white,
leftmargin=0,%
rightmargin=0,
backgroundcolor=white!10,%
innertopmargin=0pt,%
ntheorem]{sidenote}{Remark}[section]

\usepackage{array}
\newcolumntype{L}[1]{>{\raggedright\let\newline\\\arraybackslash\hspace{0pt}}m{#1}}
\newcolumntype{C}[1]{>{\centering\let\newline\\\arraybackslash\hspace{0pt}}m{#1}}
\newcolumntype{R}[1]{>{\raggedleft\let\newline\\\arraybackslash\hspace{0pt}}m{#1}}

\usepackage{tabu}

\title{Modelling instabilities in electroelastic and magnetoelastic membranes: a very brief review} % Title is tentative. Most likely it will be changed later.
\author{Prashant Saxena \\[2ex]
{\small James Watt School of Engineering,  University of Glasgow, Glasgow G12 8LT, UK}}

\date{\small This document was last updated on \today.}
\date{\small  Email: prashant.saxena@glasgow.ac.uk}

\allowdisplaybreaks % This allows equations to span over multiple pages.

\begin{document}

\maketitle

\begin{abstract}
The current state of the art for analytical and computational modelling of deformation in nonlinear electroelastic and magnetoelastic membranes is reviewed. 
A general framework and a list of methods to model large deformation and associated instabilities (wrinkling, limit point, global bifurcation) due to coupled electromechanical or magnetoemechanical loading is presented.
\end{abstract}

% \tableofcontents

% \setcounter{section}{-1}
\section{Introduction}
In this short note, I aim to summarise the techniques developed in recent years to model large deformation in electroelastic and magnetoelastic membranes.
My aim is not to perform an extensive review of the large quantity of available literature on this topic, but to provide the readers with a set of techniques and a starting point to undertake research in this area.
Membranes are defined as thin, highly stretchable structures that have negligible bending stiffness.
Membranes are typically capable of taking significant strains before onset of any plastic deformation or fracture.
Membranes made of electroelastic or magnetoelastic materials can be further actuated with electric or magnetic field, respectively.
Application of large coupled electro-mechanical or magneto-mechanical load often leads to interesting behaviour including multiple stable configurations, snap-through instability, formation of wrinkles, and loss of symmetry (global buckling).
A few papers where this has been demonstrated experimentally for electroelastic membranes are by \cite{Li2013a, Kollosche2015, Zhang2016, Mao2018}.
\cite{Raikher2008a} reported a global bifurcation from flat fundamental solution for a circular magnetoelastic membrane actuated by magnetic field.

\cite{Barham2007,Barham2008,Barham2010,Barham2012} modelled large  deformation and limit point instabilities in magnetoelastic membranes using both, analytical and finite element, methods.
\cite{Reddy2017,Reddy2018,Saxena2019} performed comprehensive analysis of all the instabilities listed above in circular, cylindrical, and toroidal magnetoelastic membranes using a variational approach.
Choice of symmetric geometries allowed them to convert the resulting partial differential equations to ordinary differential equations.
\cite{Ali2021} derived a 2D membrane theory for magnetoelasticity using asymptotic analysis and studied the deformation of an annular membrane.
\cite{Goulbourne2007} derived the equilibrium equations and studied time-dependent response of a circular inflated electroelastic membrane.
\cite{Xie2016} studied the bifurcation of a spherical electroelastic membrane to a pear-shaped configuration.
\cite{DeTommasi2011,Greaney2019} focussed on wrinkling in electroelastic membranes while \cite{Liu2021} have analysed all the instabilities listed above for a special case of inflated toroidal electroelastic membrane.
A generalisation of the tension field theory \citep{Steigmann1990} is the main tool of analysis of wrinkling instability in the research cited above.

\section{Modelling procedure}
We outline the energy based approach of modelling instabilities reported by \cite{Reddy2017,Reddy2018,Saxena2019,Liu2021}. A description of the modelling strategy is presented in the following subsections.

\subsection{Kinematics}
First we need to appropriately describe the kinematics of the membrane deformation.
The  mid-surface needs to be modelled as a two-dimensional surface but the thickness stretch ratio is also an important kinematic quantity that requires consideration in this problem.
As an example, if we consider the toroidal membrane described in \citep{Reddy2017,Venkata2020, Liu2021}, we need the functions $\varrho$ and $\eta$ to describe the deformed mid-surface.
We write the metric tensors $\mbf{G}$ and $ \mbf{g}$ in the local curvilinear coordinate system in the reference and deformed configurations, respectively, and use them to compute the deformation gradient tensor $\Dgrad$.
On application of the constraint of incompressibility $(\text{det} (\Dgrad) = 1)$, we can write the through thickness stretch ratio $\lambda_3$ in terms of the functions $\varrho$ and $\eta$.

\subsection{Total energy of the system}
Next, we need to write the total energy of the system that needs to be minimised to determine the equilibrium equations and stability conditions.
The total energy in this case consists of the electroelastic/ magnetoelastic energy stored in the membrane material due to deformation, the energy of the surrounding electric/ magnetic field, and the work done due to the applied pressure. 
A comprehensive description of these energies appropriate to the problem and procedure of their minimisation is given by \cite{Saxena2020, Sharma2021} and in the book by \cite[Chapter 8]{Dorfmann2014b}.
The energy formulations cited above are derived for three-dimensional continua and they need to be simplified for use in the current context.

Electroelastic membranes are typically made of dielectric elastomers such as VHB. Conductive grease is applied to thin sheets of this elastomer \citep{Li2013a, Kollosche2015}. This configuration allows application of a potential difference across the thickness of the membrane that localises the electric field within the membrane and the electric field in the surrounding free space can be neglected (see \citep[Section 3.1]{Liu2021}).
Total energy for thin magnetoelastic membranes can be simplified using the weak magnetisation assumption.
We allow only one-way coupling, that is, the magnetic field causes a deformation of the membrane but the deformation in the membrane is not strong enough to significantly perturb the surrounding magnetic field.
Thus the magnetic field in the free-space can essentially be considered as constant (see \citep{Barham2007} and \citep[Section 2.2.1]{Reddy2017}).
These assumptions can drastically simplify the problem and reduce the number of unknowns.

Vanishing of the first variation of the energy results in the governing equilibrium equations for the problem.
For the equilibrium to be stable, the solution should be a minimiser of the energy.
This is checked by computation of the second, and sometimes, higher variations. Details of this are described later.

\subsection{Limit point or snap through instability}
% The equilibrium equations derived from the first variation can be solved using standard numerical methods.
% For certain special geometries, partial differential equations can be converted to ordinary differential equations.

It is typically observed that in a pressure-controlled inflation of membranes, volume of the enclosed fluid can suddenly increased drastically at some critical value of pressure, see Figure \ref{fig: wrinkles}a.
This sudden change is called as snap-through and results in significant change in the membrane strains.
It is also often referred to as the limit point instability.
Solution of the equilibrium equations derived from the first variation can clearly demonstrate a peak in the pressure-volume curve (see \citep[Figure 4]{Liu2021}).
In order to accurately capture the behaviour of the system post the limit point, appropriate arc-length procedure needs to be adopted in the numerical scheme to accurately determine all the equilibrium states.
A general procedure is given by \cite{Kadapa2021}.
A simplified procedure for solving ODEs while following the curve can be seen in \citep[Section 2.4]{Reddy2017}.

\subsection{Wrinkling instability}
\begin{figure}
 \centering
 \begin{tabular}{c c}
 \includegraphics[width=0.45\linewidth]{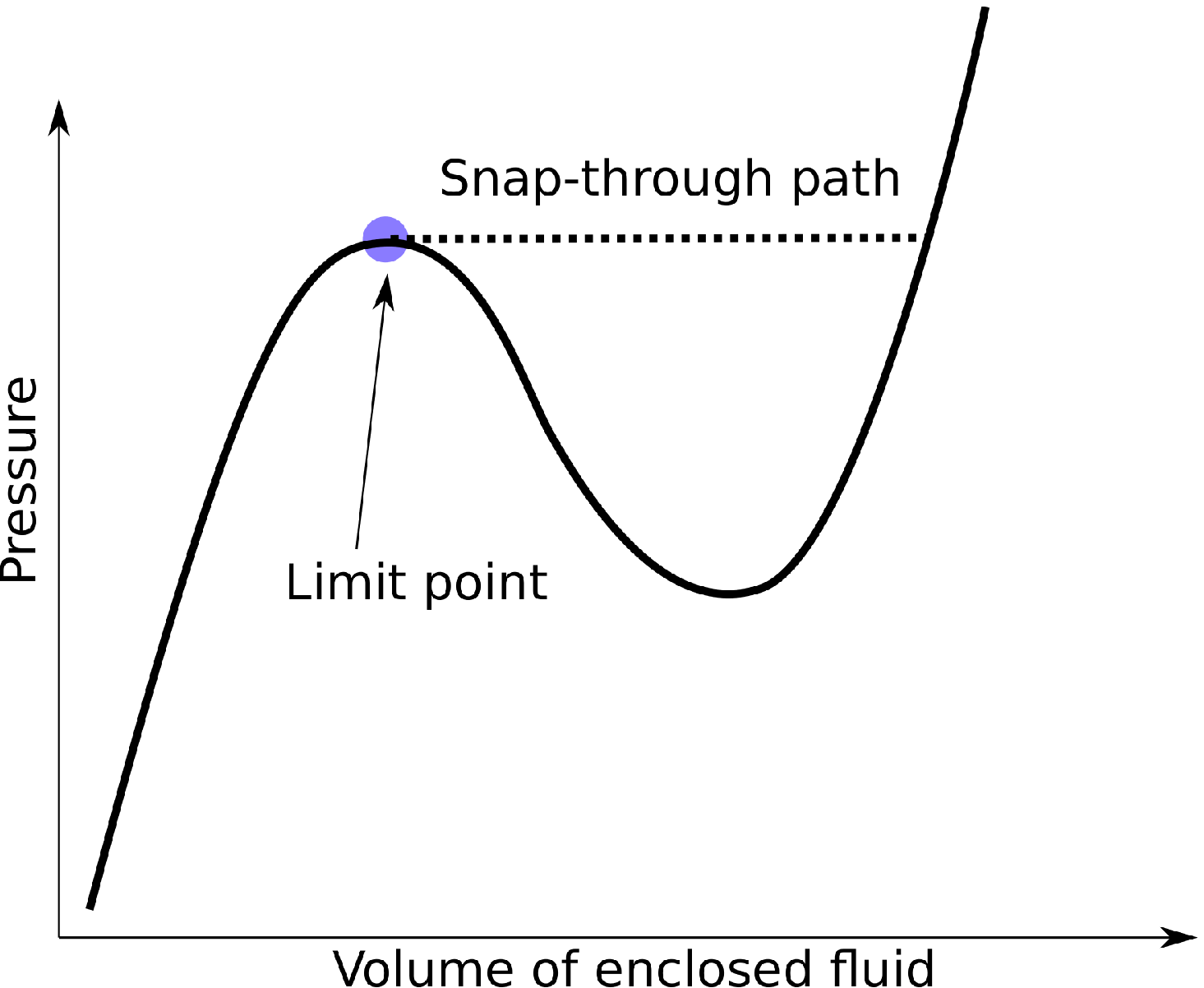}& \includegraphics[width=0.4\linewidth]{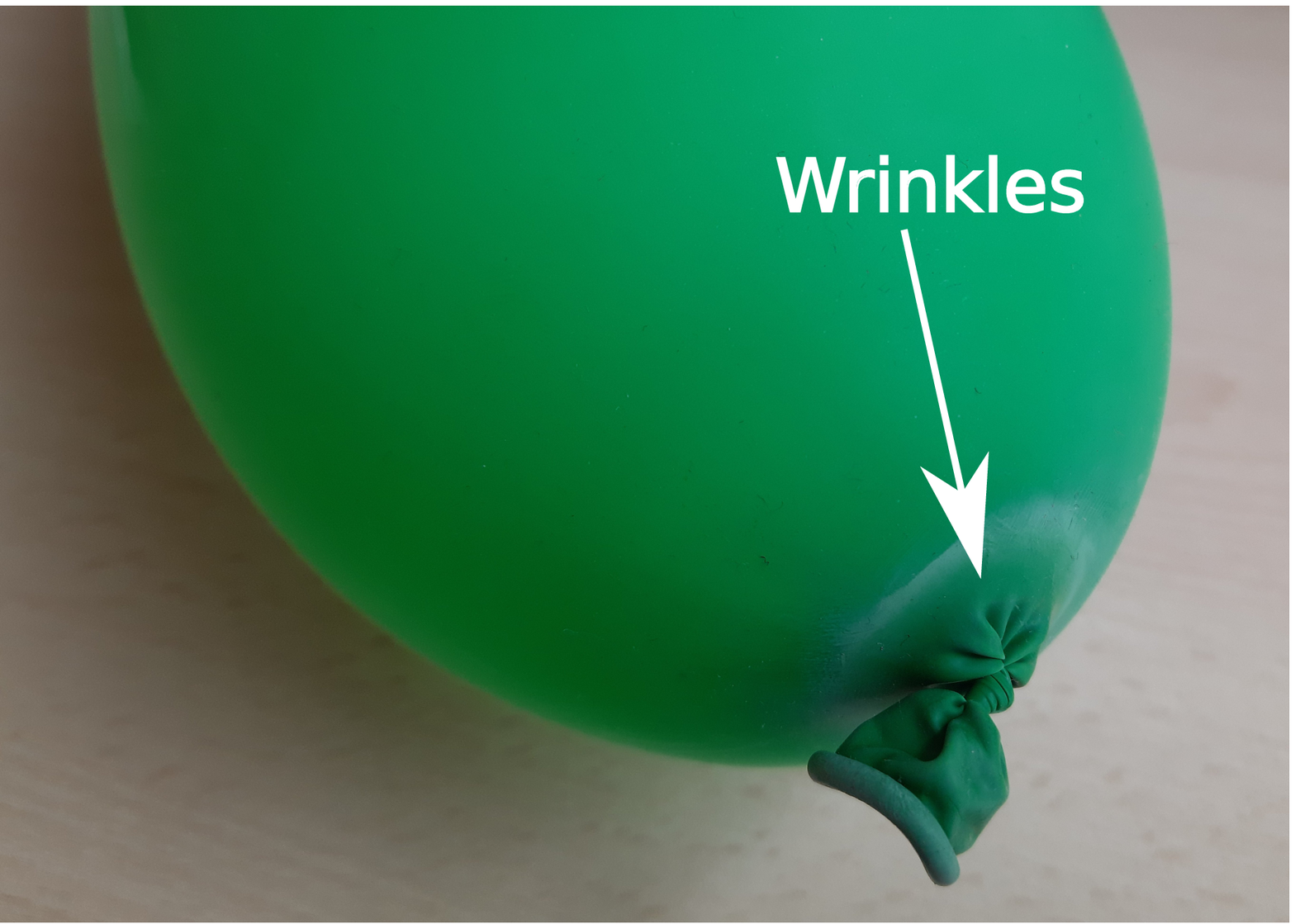} \\
 (a) & (b)
 \end{tabular}
 \caption{(a) A depiction of the typical snap-through or limit point instability in inflated membranes. (b) Formation of wrinkles in an inflated party balloon at the location of compressive stresses.}
 \label{fig: wrinkles}
\end{figure}

An ideal membrane is assumed to have zero bending stiffness and can sustain only tensile loads. 
It offers no resistance to in-plane compression, instead it leads to formation of wrinkles to relieve the stress, see Figure \ref{fig: wrinkles}b.
If the solution of equilibrium equations is a deformation state that results in compressive stresses, then that solution is likely to be physically incorrect.
One needs to reformulate the equations using the tension field theory \citep{Steigmann1990} by replacing the strain energy density function with  a relaxed strain energy density function.
According to the tension field theory, as the principal stress in one direction approaches to zero from a tensile state, wrinkles form orientated perpendicular to this direction. 
Due to an assumption of zero bending stiffness, this theory is not able to quantify the frequency or amplitude of wrinkles but outputs the critical load for onset of wrinkles and the direction of resulting wrinkles.
An extension of the tension field theory to the case of electroelasticity is shown by \cite{DeTommasi2011, Greaney2019, Liu2021}. 
\cite{Saxena2019} have simply checked the in-plane stress condition to predict wrinkling for magnetoelastic membranes but haven't recomputed the updated solution.
% and to the case of magnetoelasticity by \cite[Section 3.4]{Saxena2019}.

\subsection{Global bifurcation}
\begin{figure}
 \centering
 \includegraphics[width=0.5\linewidth]{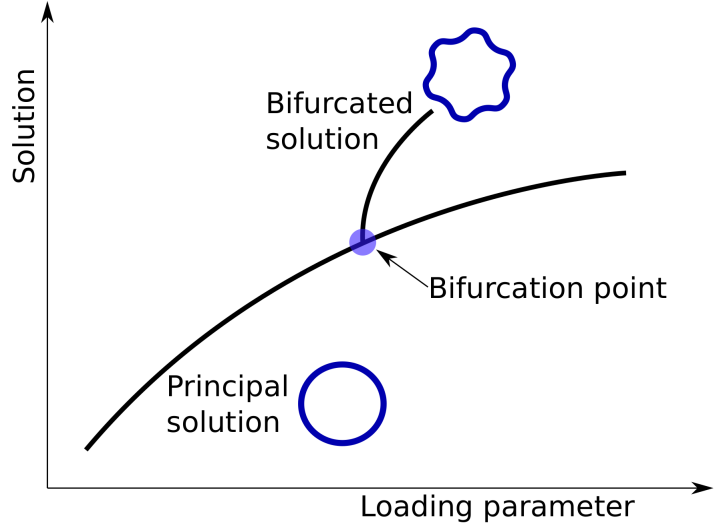}
 \caption{Bifurcation of solution from the circular principal solution to an oscillatory solution superposed on a circle.}
 \label{fig: bifurcation}
\end{figure}
Wrinkles in membranes are localised instabilities that form to relieve compressive stresses.
The membrane can also undergo a global instability that leads to change in the overall solution as schematically depicted in Figure \ref{fig: bifurcation}.
There are two main ways to model such bifurcation phenomena.

We can use the theory of incremental deformations superimposed on a finite deformation to determine the critical point of onset of this bifurcation \cite[Chapter 6]{Ogden1997b}.
Typically this involves determining the equilibrium equations for the bifurcated solution by setting the second variation of the energy functional to zero.
One can then derive the conditions that need to be satisfied for the bifurcated solution to exist.
Computation of this condition leads to bifurcation point (the value of load at which the principal solution is no longer a minimiser of the energy since the second variation is no longer positive).
For an application of this procedure to electroelastic membranes, see \cite[Section 5]{Liu2021}.
It is worth noting that this procedure, when used in practice, can inform only about the type of bifurcation that one looks for.
For example, in the case of surface instabilities in bulk elastic solids, it is well known that the theory of incremental deformations doesn't always give the correct bifurcation point \citep{Hong2009,Hohlfeld2011}.

An alternative method is to directly write the second variation of the energy functional as
\begin{equation}
 \delta^2 E = \int\limits_{\Omega} \big[ \langle \bsym{P \Delta}', \bsym{\Delta}' \rangle + \langle \bsym{Q \Delta}, \bsym{\Delta} \rangle \big] d \mbf{x} , \nonumber
\end{equation}
where $\bsym{\Delta}$ is the perturbation in the solution, and $\bsym{P}$ and $\bsym{Q}$ are matrices. 
One can then define  necessary and sufficient conditions based on certain theorems in calculus of variations and check for stability of solution, see \cite[Section 2.5]{Reddy2017} and \cite[Chapter 5]{Gelfand2003}.
An advantage of this method over the previous one is that it directly looks at the second variation instead of looking for bifurcations of a certain type.
It can also distinguish between solutions that are definitely stable, definitely unstable, and ``probably stable''. 
The third type of solution (probably stable) correspond to the case in which the necessary condition of stability is satisfied but the sufficient condition is not, see \cite[Figure 11]{Reddy2017}.
However, it is effective only as a check of stability and doesn't given any information on the direction of perturbation.
A combination of the above two methods can deliver a more comprehensive understanding of the bifurcation point.

\subsection{Post bifurcation analysis and some outstanding issues}
Little work has been done on analysing the deformation of membranes beyond the global bifurcation. 
This is  an area that provides a great deal of opportunity for new research on this topic.
\cite{Xie2016} have analysed the post-buckling deformation of a spherical electroelastic shell that bifurcates to a pear-shaped configuration.
Spherical symmetry significantly simplifies the equations and they use the results for a purely elastic spherical shell \citep{Haughton1978} as a guiding condition for their bifurcation analysis.
In general, one can follow Koiter's method to compute the post-bifurcation response of inflated membranes \citep{Koiter1945,Budiansky1974}.

The wrinkling instability computed using the tension field theory \citep{Steigmann1990} is currently based on an ad-hoc extension to electroelasticity by \citep{Greaney2019}.
A mathematical rigorous derivation of magnetoelastic or electroelastic tension field theory is still an open problem. Computation of the wavelength and amplitude  of wrinkles would require a further generalisation of the membrane approximation to nonlinear shell approximation that could account for resistance to bending.

% \newpage
\bibliographystyle{./author-year-prashant}
\bibliography{/home/prashant/Documents/library}
% \bibliography{./references_used2}
\end{document}